\documentclass[12pt]{iopart}
\usepackage{cite}
\usepackage{epsf}
\usepackage{epsfig}
\usepackage{float}
\usepackage{amssymb}
\usepackage{amsfonts}
\usepackage{color}

\def\P{{\bf{P}}}

\def\beq{\begin{equation}}
\def\eeq{\end{equation}}
\def\beqa{\begin{eqnarray}}
\def\eeqa{\end{eqnarray}}

\newcommand\bear{\begin{eqnarray}}
\newcommand\eear{\end{eqnarray}}
\newcommand\bea{\begin{align}}
\newcommand\ena{\end{align}}
\usepackage{comment}
\usepackage{setspace}

\begin{document}

\title[Breakdown of electron-pairs in presence of electric field of a superconducting ring]{Breakdown of electron-pairs in presence of electric field of a superconducting ring}

\author{ Bradraj Pandey$^{1}$, Sudipta Dutta$^{2}$, and Swapan K. Pati$^{1}$}
\address{$^{1}$Theoretical Sciences Unit Jawaharlal Nehru Centre for Advanced Scientific Research,\\
 Jakkur, Bangalore 560 064, India.\\ $^{2}$
Indian Institute of Science Education and Research (IISER),
Tirupati, Rami Reddy Nagar, Karkambadi Road, Mangalam B. O., Tirupati 517 507, India}

\begin{abstract}
The quantum dynamics of quasi-one-dimensional ring with varying electron filling 
factor is investigated in presence of external electric field. The system is 
modeled within Hubbard Hamiltonian with attractive Coulomb correlation, which
results in superconducting ground state when away from half-filling. The electric 
field is induced by applying time-dependent Aharonov-Bohm flux in the perpendicular 
direction. To explore the non-equilibrium phenomena arising from the field, we adopt 
exact diagonalization and Crank-Nicolson numerical method. With increase in electric
field strength, the electron pairs, a signature of superconducting phase, start
breaking and the system enters into a metallic phase. However, the strength of the
electric field for this quantum phase transition depends on the electronic
correlation. This phenomenon has been confirmed by flux-quantization of
time-dependent current and pair correlation functions.
\end{abstract}


\maketitle

\section{Introduction}
The strongly correlated low-dimensional systems and their response to the
external perturbations, e.g., applied field has been an ever-growing research
area owing to their rich quantum phase diagram. Recent advancements in 
experiments enable the realization of the quantum dynamics in 
materials\cite{iwai,okamoto,swano,taguchi} and in the cold atom 
system\cite{bloch,hoff,greiner,kohl} in non-equilibrium environment.
Dielectric breakdown of Mott insulating phase in organic thyristor\cite{swano} 
and one-dimensional systems, such as, $Sr_2CuO_3$ and $SrCuO_2$\cite{taguchi}
has been realized experimentally in presence of strong electric field. There has
been report of photo-induced Mott transition in halogen-bridged Ni-chain compound
as well. In cold atom systems, such as, one-dimensional Bose gases, the 
non-equilibrium dynamics of superfluids has been studied \cite{hoff}. 
Three dimensional fermionic optical lattices also exhibit band-insulator
to metal transition by controlling the interaction between atoms through the 
Feshbach resonance\cite{kohl}.

In a recent experiment, breaking of electron-pairs, namely Cooper-pairs
of superconducting system is shown by exposing it to photon flux\cite{pjde}.
At low enough temperature, superconductors are condensate of Cooper-pairs
which are sensitive towards the external perturbations. Motivated by these
experiments, here we study the breaking of electron pairs in superconducting
rings, which can be realized experimentally \cite{filby,upfer} along with
the measurement of their persistent current\cite{levy,deblock,kosh}. The
superconducting rings are described by the attractive Hubbard model.
Here the net effective attractive  interaction forms the electron pairs,
leading to superconductivity\cite{penc,ferr,kusa}.
 In the real materials, the origin of this attractive interaction
 can also be due to the coupling between electron and lattice, excitons or plasmons
\cite{rmic}. For lower values of attractive interaction, electrons form BCS (Bardeen-Cooper-Schrieffer)
type of pairing with loosely bound pairs, while increase in attractive interaction
results in strong local pairing (BEC-limit (Bose-Einstein condensate))\cite{mohit}.
These strong-pairs that are similar to charged bosons
can condensate and give rise to superconducting state\cite{rmic}.
Therefore, superconductivity requires formation of electron pairs and
phase-coherence between the pairs.

External electric field induces fluctuation in these pairs and eventually breaks them.
We have modeled the external electric field in terms of time-dependent Aharonov-Bohm (AB) flux
and studied  the non-equilibrium properties and time evolution
of many body wave function, by using exact diagonalization and Crank-Nicolson
method. We found that the breaking of electron-pairs depends on the strength of
electric field and attractive interaction.
We analyze this depairing of electrons by flux-quantization of
persistent-current, time average current and pair-correlation function.

The flux quantization\cite{onsag,byers} and the off-diagonal long-range order\cite{yang}
 are the key characteristics of superconductors.
 The quantization of AB-flux has been observed experimentally in conventional\cite{doll,bsd}
 and in the high-temperature superconductors\cite{ceg}.
For the free electron system, the quantized flux is $\phi_0=hc/e$\cite{ab}.
 In case of superconductors, the electron-pairing results in
halved period, i.e., $\phi = \phi_0/2$  \cite{onsag,byers,bbsch,sudbo,cahay,ased}.
 Similar phenomenon can be observed in case of repulsive Hubbard model
 for finite size rings due to the spin degrees of freedom\cite{lchen},
 even without superconductivity. This ambiguity has driven us to choose
the more global approach of extended-AB period method to detect the
electron-pairing\cite{kusa,arita,kusa2}, since this keeps track of the evolution
of energy levels and the wavefunction as a function of the flux over the extended-AB
period $(0 \le \phi \le L \phi_0)$ . In this approach, the quantized flux for the free
electron system has been considered to be $L \phi_0$, where $L$ is the total length of
the system. However, electron-pairing in superconducting state makes this
extended-AB period halved, i.e., $L\phi_0/2$ (for more detail about extended-AB period 
we refer to \cite{kusa2}). But whenever there is a formation of the density wave 
(charge or spin), this periodicity gets confined within the lattice constant, i.e., $\phi_0$.

\section{Model and numerical method}

%******************************** Fig.1 *****************************
\begin{figure}[ht]
\begin{center}
\includegraphics[width=2.8in]{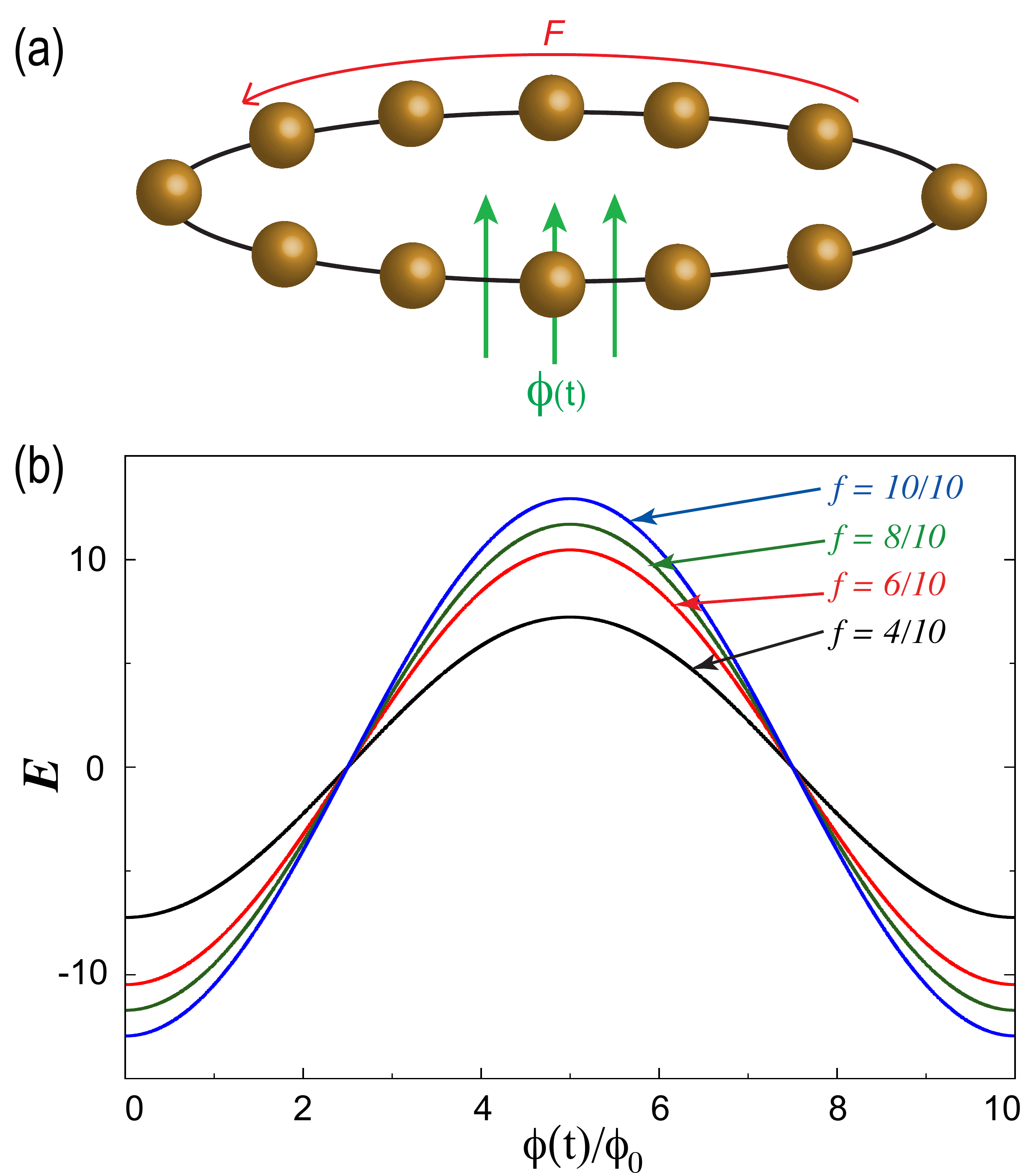}
\caption{(a) The schematic of quasi-one-dimensional ring with 12 sites.
The time dependent perpendicular AB flux, $\phi(t)$ generates the
circulating electric field, $F$ in the ring.
(b) Time evolution of the non-interacting ($U = 0$) ground state energy $E$,
as a function of $\phi(t)/\phi_{0}$, for $F=0.0005$ with varying filling factors $f$.}
\end{center}
\end{figure}
%*******************************************************************

To investigate the behavior of these electron-pairs in presence of external
electric field, we consider the quasi-one-dimensional ring structure
(see Fig.1(a))  and modeled the system within
attractive Hubbard model,

\begin{equation}
H(t) = -\gamma\sum_{i,\sigma} \left(\exp^{2 \pi i\phi(t)/N}c^{\dagger}_{i+1,\sigma}c_{i,\sigma} +h.c\right)
-U \sum_i n_{i,\uparrow}n_{i,\downarrow}
\end{equation}

\noindent where $\gamma$ and $U$ are the hopping term and the attractive onsite
Coulomb potential. $c^{\dagger}_{i,\sigma}$ ($c_{i,\sigma}$) creates (annihilates)
one electron with spin $\sigma$ at $i$-th site and $n$ is the number operator.
The electric field $F$ has been included in terms of time-dependent AB-flux,
$\phi(t)=eFLt$ (see Fig.1(a)),  $N=L/a$ denoting the number of sites, $a$ is
lattice constant  and $t$ being the time. We set $e=h=a=1$ and
assume $\gamma$ as the unit of energy throughout the paper. Since, $a=1$, $L$
has been considered to be same as $N$ in the following sections.

We consider different system lengths ($L$ = 8, 10 and 12), with various filling factors $f$
($f=n_e/L$), where $n_e$ is the number of electrons in the system). Note that, we always 
consider same number of up and down spins to keep the z-component of the total spin, $s^{total}_{z}$
as zero. In absence of electric field, the above model results in superconducting ground 
state\cite{psch,rmic,penc,ferr,kusa}, while away from half-filling. However, at half-filling 
the superconducting state becomes degenerate with a charge density wave ground state. 
At half-filling, a large enough negative value of $|U|$  
ensures the pairing of electrons at alternate sites. Therefore, 
the ground state becomes alternate bound pairs and empty sites,
which constitutes a charge density wave ($|\uparrow \downarrow, 0, \uparrow \downarrow,0, 
\uparrow \downarrow,0... \rangle$)\cite{afho}. For finite values of $U$, the system 
has paring gap of the order of $U$, to spin excitation. The degeneracy can be lifted by 
deviation from half-filling, which essentially tilts the stability towards superconducting 
phase \cite{rmic}. Note that, at half-filling, the repulsive Hubbard Hamiltonian, on the contrary, 
leads to a Mott insulating phase, characterized by antiferromagnetic spin density wave to avoid 
on-site pairing of electrons. Unlike the attractive $U$ case, here the spin sector becomes gapless,
while opening up the charge gap.

We adopt exact diagonalization method to solve the above Hamiltonian
and to obtain ground sate wave function, $|\psi(0)\rangle$, at $t=0$.
Then we evolve $|\psi(0)\rangle$ with time by solving the time-dependent
Schrodinger equation: $i\frac{d}{dt}|\psi(t)\rangle = H(t)|\psi(t)\rangle$.
For the time evolution of $|\psi(t)\rangle$ at  absolute zero temperature,
we adopt Crank-Nicolson's algorithm which preserves the unitary time
evolution without divergence at large time limit. The time evolution can
be written as,

\begin{equation}
|\psi(t+\delta t)\rangle = \exp^{-i \int^{t+\delta t}_t H(t) dt} |\psi(t)\rangle
 \simeq \frac{1-i\frac{\delta t}{2}H(t+\frac{\delta t}{2})}
{1+i\frac{\delta t}{2}H(t+\frac{\delta t}{2})}|\psi(t)\rangle
\end{equation}

For precise convergence of the wave-function, the time step $\delta t$ has
been considered to be small enough, 0.01 in units of $\hbar/\gamma$. This unit
of time has been chosen to make the exponential dimensionless.
Note that, each time evolution step requires the computationally expensive matrix
inversion of the many body Hamiltonian matrix. We adopt Davidson algorithm
for this purpose, which gives proper convergence.

\section{Results and discussion}

%********************************************************************

First we investigate the flux quantization of the ground state energy 
and the current density as a response to the applied AB-flux with 
varying filling factor, $f$ and attractive potential. Then the effect
 of electric field on the ground state has been studied to detect the 
superconducting to metallic phase transition. This has been characterized
by further calculations on pair correlation function and the time averaged current.

%******************************** Fig.2 *****************************
\begin{figure}[ht]
\begin{center}
\includegraphics[width=3.4in]{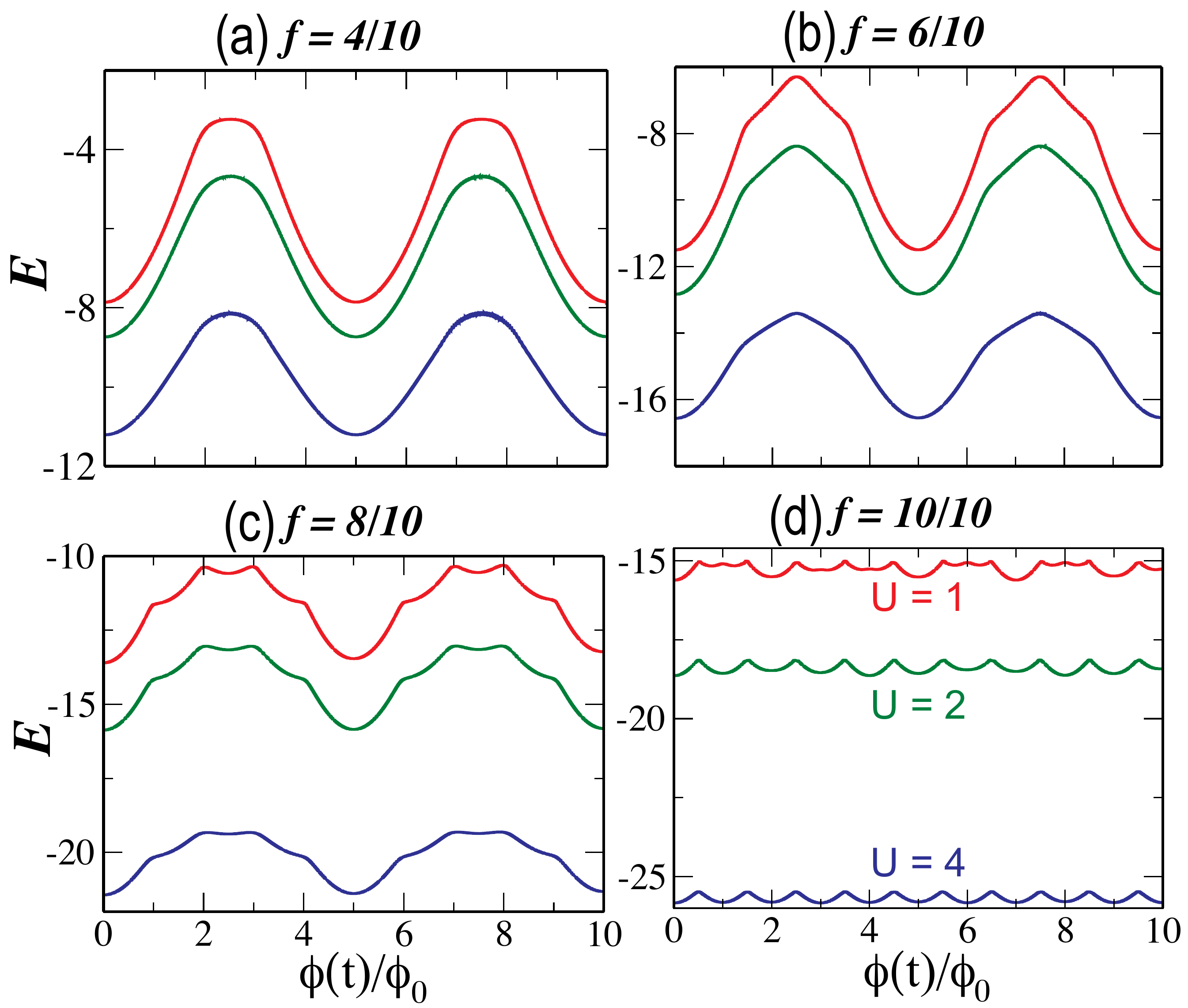}
\caption{The time evolution of the interacting ($U \ne 0$) ground state energy $E$,
 as a function of $\phi(t)/\phi_{0}$, for $F=0.0005$ with varying $U$ for
(a) f =4/10, (b) f =6/10, (c) f =8/10 and (d) f =10/10.}
\end{center}
\end{figure}
%*******************************************************************

%******************************** Fig.3 *****************************
\begin{figure}[ht]
\begin{center}
\includegraphics[width=3.4in]{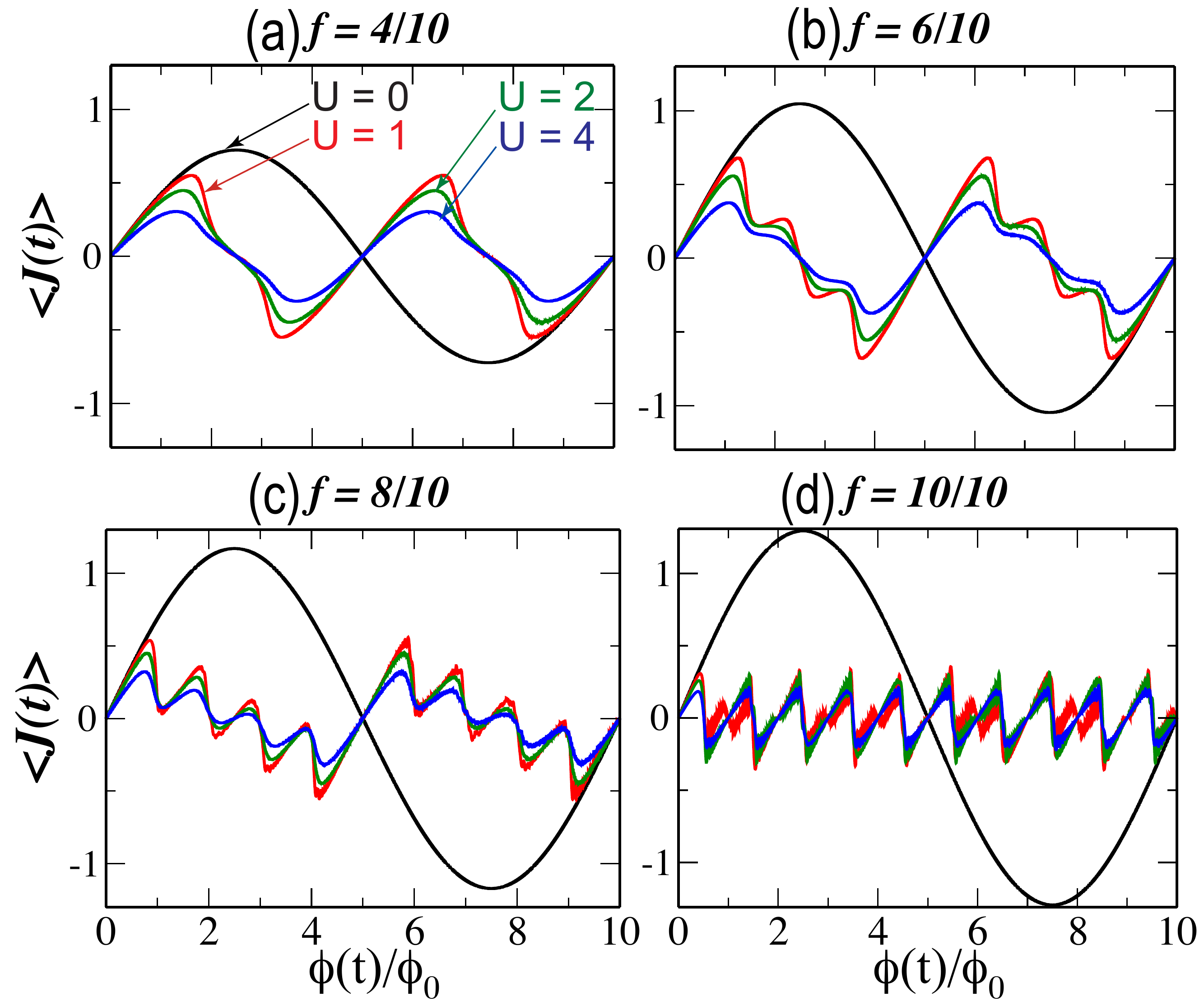}
\caption{Time evolution of current-density, $\langle J(t)\rangle$, as a function of $\phi(t)/\phi_{0}$
 for  $F=0.0005$, with different values of $U$ and filling factors
 (a) $f$ =4/10, (b) f =6/10, (c) f =8/10 and (d) f =10/10.}
\end{center}
\end{figure}
%********************************************************************

In Fig.2, we show the time evolution of the ground state energy for interacting case,
as a function of AB-flux for very small $F$.
As can be seen, for non-interacting systems (Fig.1(b)),
the periodicity is always equal to the extended AB-period, $L\phi_0$.
Once the attractive interaction is turned on, the periodicity becomes halved, i.e., $L\phi_0/2$
in case of smaller $f$ values (Fig.2(a)), indicating superconducting ground
state with the formation of electron-pairs. However, increase in $f$ results
in additional cusps, coexisting with the halved extended AB-period (Fig.2(b) and (c)).
These cusps are signature of level anti-crossings, arising from the enhanced degeneracy
in the system with higher number of electrons\cite{kusa,kusa2}. These degenerate states are connected
via two-particle scattering processes along with the Umklapp processes \cite{kusa2}.
Once the system attains half-filling ($f=1$), the periodicity reduces to $\phi_0$ (Fig.2(d)),
as the system forms charge density wave phase, with pairing up of two electrons 
with opposite spins at alternate sites under the influence of attractive $U$\cite{kusa2,afho}.

For further characterization of flux quantization,
we investigate the current density as a function of the applied AB-flux.
The current density operator is defined as.
\begin{equation} 
J(t) = -\gamma \sum_{i,\sigma}i \left(\exp^{2 \pi i\phi(t)/N}c^{\dagger}_{i+1,\sigma}c_{i,\sigma} -h.c. \right)
\end{equation}

In Fig.3, we show the time evolution of current density as a function of $\phi(t)=FLt$.
Note that the current density is the change in the slope of energy with respect to the flux.
Therefore, one can find direct correspondence between the energy and
the current density plots as a function of flux (see Fig.2 and Fig.3).
As expected, the $<J(t)>$ shows extended AB-period for non-interacting system.
The fractional periodicity\cite{fvkusma, ferrari} in the current (see Fig.3(b) and (c)) corresponds
to the cusps of energy. The coexistence of superconducting state,
characterized by the halved period and the fractional periodicity can also
be seen from $<J(t)>$ plot for $f<1$. Note that, due to the finiteness of
the one-dimensional system, the kinetic energy of the electron-pairs tries
to break the pairing and leads to the formation of fractional periodicity.
Therefore, increase in system length or the increase in attractive potential,
$U$ can lead to stable superconducting ground state with improved half-periodicity\cite{kusma}.
This behavior is clearly visible from $U$ dependence of current density in Fig.3.

%******************************** Fig.4 *****************************
\begin{figure}[ht]
\begin{center}
\includegraphics[width=4.0in]{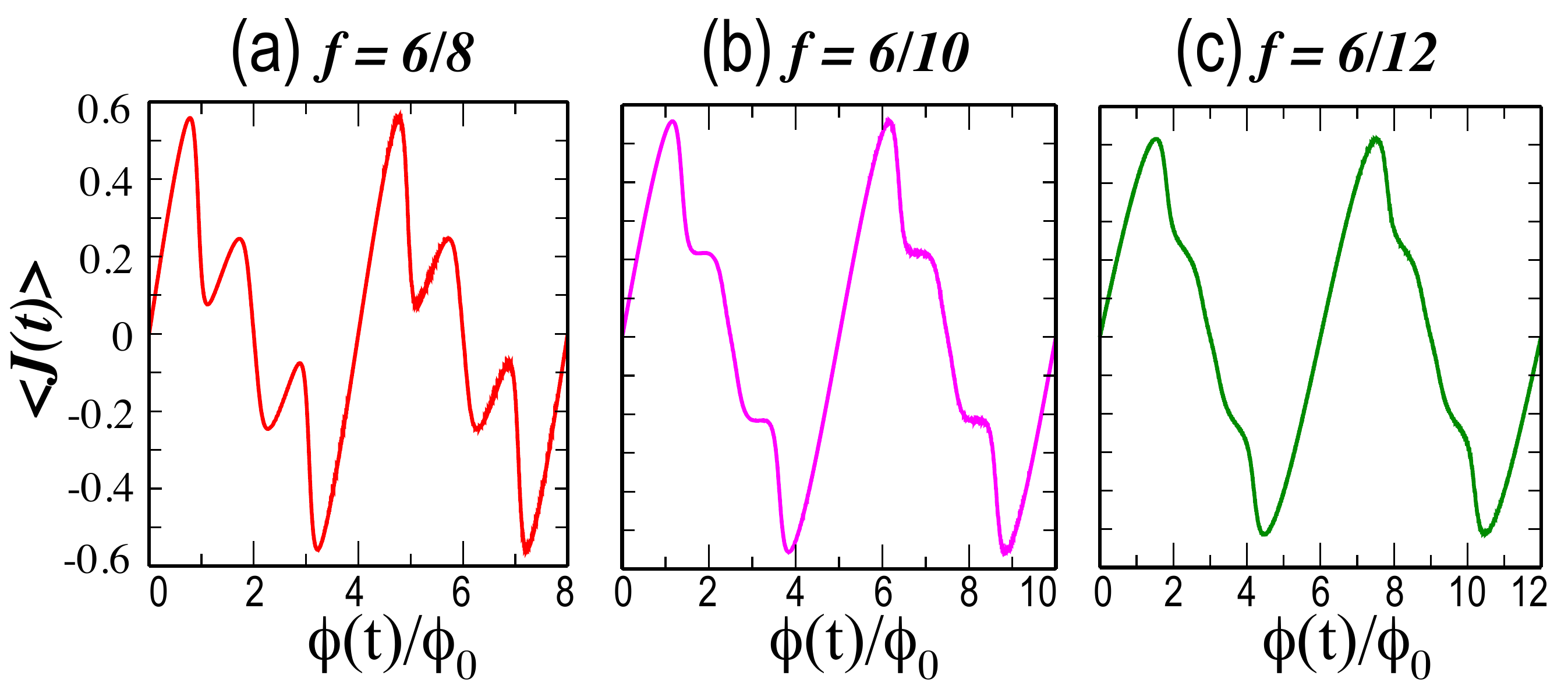}
\caption{Time evolution of current-density, $\langle J(t)\rangle$ as a function of $\phi(t)/\phi_{0}$
 for  $F=0.0005$, $U = 2$, and different system sizes (a) $L=8$, 
(b) $L=10$, and (c) $L=12$ with fixed number of electrons $n_e=6$.}
\end{center}
\end{figure}
%********************************************************************

Furthermore, we systematically increase the system size for a fixed number of electrons,
$n_e = 6$ and fixed $U = 2.0$ and show current density in Fig.4.
As can be seen, with gradual increase in system size from 8 to 10 to 12,
the halved periodicity becomes smoother and the fractional periodicities gradually disappear\cite{kusma}.

%******************************** Fig.5 *****************************
\begin{figure}[ht]
\begin{center}
\includegraphics[width=3.4in]{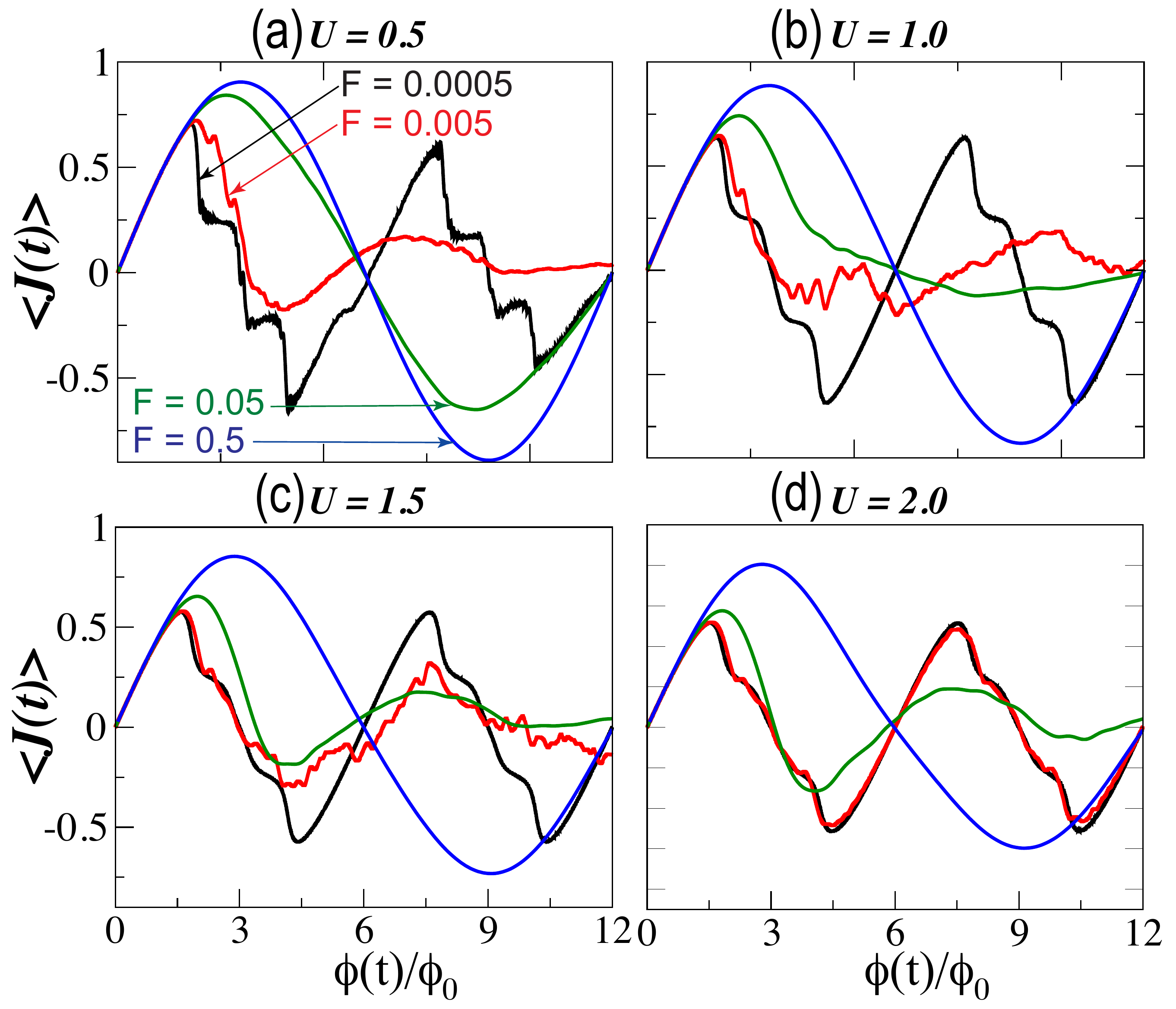}
\caption{Time evolution of current-density, $\langle J(t)\rangle$
 as a function of $\phi(t)/\phi_{0}$ with varying electric field strength, $F$
for different values of attractive interaction, (a)$U$ = 0.5,
(b) $U$ = 1.0, (c) $U$ = 1.5 and (d) $U$ = 2.0.}
\end{center}
\end{figure}
%*************************************************************

Next, we investigate the effect of electric field on the ground state of superconducting
ring consisting 12 sites  with filling factors $f = 1/3$ and $1/2$.
In Fig.5, we show the current density as a function of $\phi(t)=FLt$ for varying attractive potential,
$U$ and strength of electric field, $F$. As can be seen in Fig.5(a), the superconducting
ground state (characterized by halved extended AB-period) undergoes the transition to
metallic phase (characterized by the extended AB-period) even in case of weak electric field.
This is due to fact that, the presence of weak $U$ forms loosely bound electron pairs which can
easily be broken. However, the gradual increase in $U$ strengthens the electron-pairs and the system
requires stronger electric field for the superconducting to metallic phase transition (see Fig. 5(b), (c) and (d)).
The fractional periodicities also disappear, owing to the fact that, the applied electric
field closes up the gaps at level anti-crossings, allowing the spectral flow.

For further characterization of this phase transition, we investigate the 
time evolution of the pair-correlation function as follows\cite{albert},

\begin{equation}
P(r)=\langle c^{\dagger}_{1,\uparrow} c^{\dagger}_{1,\downarrow} c_{r,\downarrow}
c_{r,\uparrow}\rangle - \langle c^{\dagger}_{1,\uparrow}c_{r,\uparrow}\rangle
 \langle c^{\dagger}_{1,\downarrow}c_{r,\downarrow}\rangle
\end{equation}

%******************************** Fig.6 *****************************
\begin{figure}[ht]
\begin{center}
\includegraphics[width=3.3in]{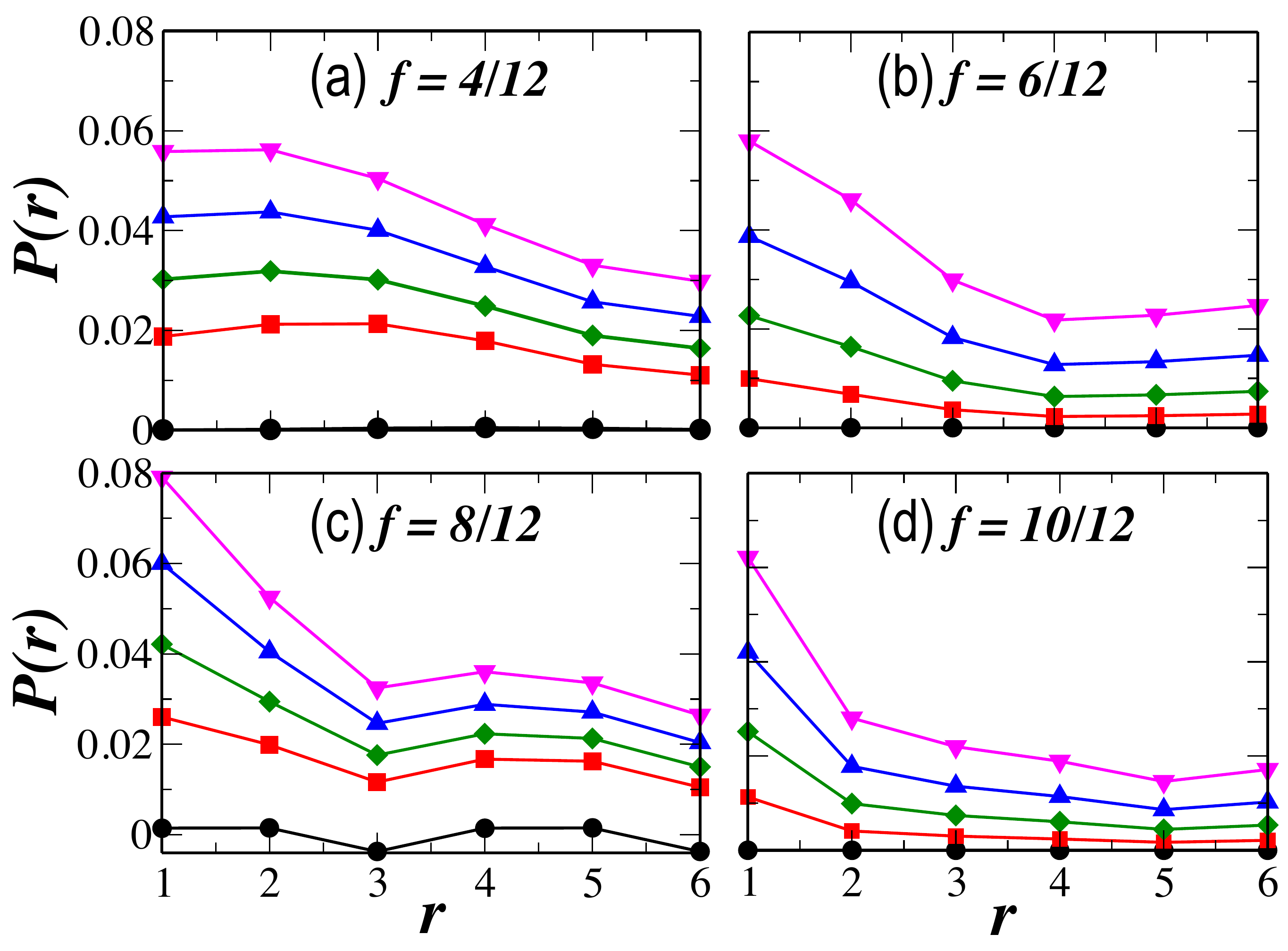}
\caption{  Pair correlation $P(r)$ as a function of $r$,
for a quasi-one-dimensional ring with $N = 12$ and 
with different attractive interaction ($U=0.0$ (circle), $U = 0.5$ (square),
$U = 1.0$ (diamond), $U = 1.5$ (upper-triangle), $U = 2.0$ (lower-triangle)) for 
different filling factors (a)$f=4/12$, (b) $f=6/12$, (c)$f=8/12$, and (d)$f=10/12$.} 
\end{center}
\end{figure}
%********************************************************************

%******************************** Fig.7 *****************************
\begin{figure}[ht]
\begin{center}
\includegraphics[width=4.0in]{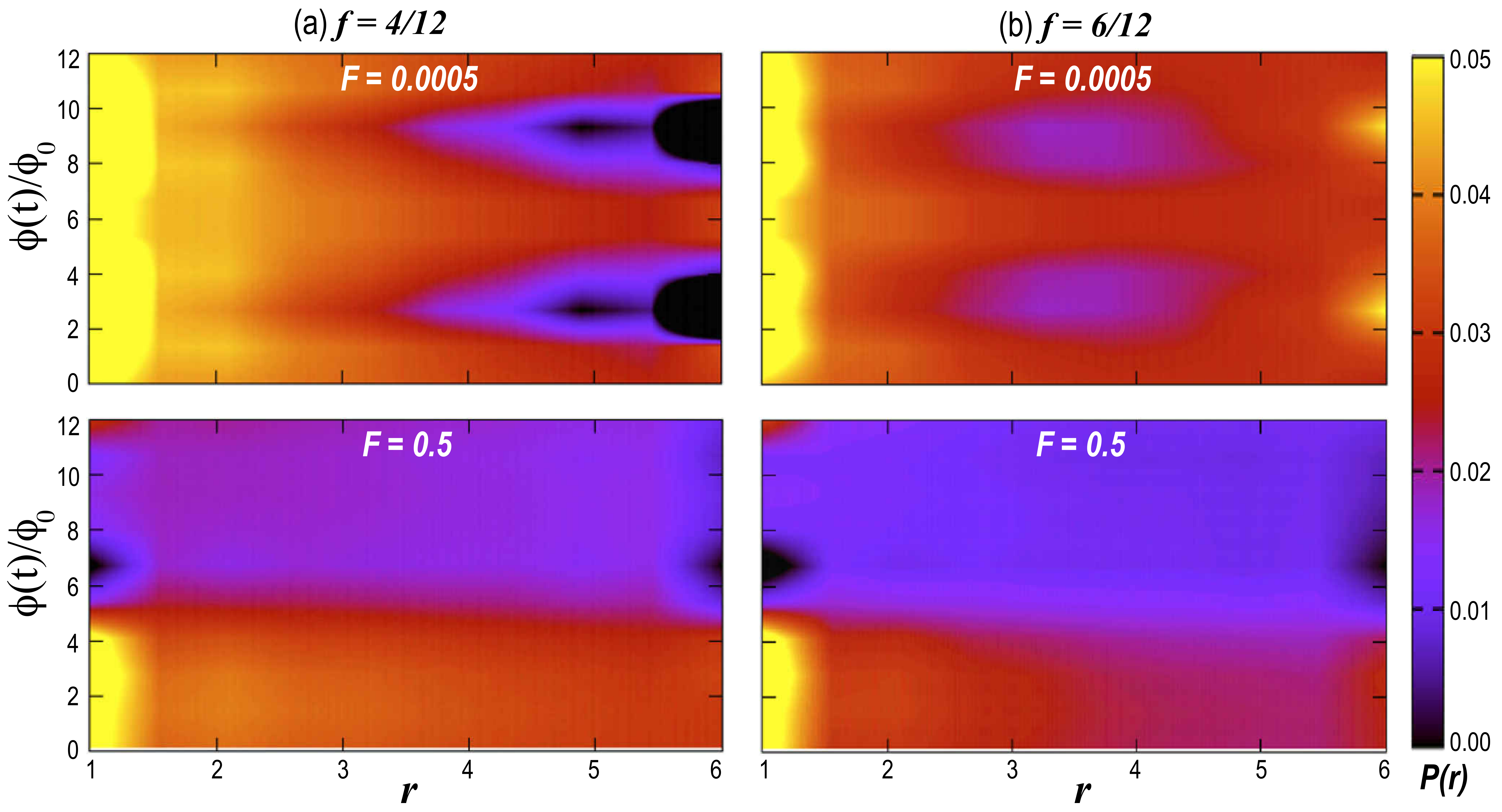}
\caption{Contour plot of pair correlation as a function of $\phi(t)/\phi_{0}$ and distance $r$,
 for the 12 site ring with $U= 2.0$ and filling factors (a)$f= 4/12$ and (b)
 $f=6/12$ with electric field strengths $F=0.0005$ (top panel) and $0.5$ (bottom panel). 
The color bar represents the numerical values of $\langle P(r)\rangle$.}
\end{center}
\end{figure}
%********************************************************************

First we have calculated pair-correlation function $P(r)$, for static case $(\phi=0.0)$.
In Fig.6 we have shown plot of  $P(r)$ as a function of $r$ for different values of $U$
 and different filing factors.
We have calculated $P(r)$ for 12 sites fermionic ring, but due to periodic boundary condition we have shown P(r)
only for 6 sites.
It is known that, the quasi long-range order of the pair-correlation function is indicative of
the superconducting phase at low-dimension \cite{albert}.
 As shown in the Fig.4, values of $P(r)$ increase with attractive interaction $U$,
which indicates the formation of strong electron bound pairs and
increase in phase-coherence within pairs.
On the other hand, in metallic case it takes either zero or very small nonzero values.

We present the expectation value of pair-correlation, $\langle P(r)\rangle$
as a function of distance $r$ and AB flux, $\phi(t)$ in Fig.7 for two different
electric field strengths, e.g., $F$ = $0.0005$ and $0.5$ and for different filling factors. In presence of
weak electric field, $\langle P(r)\rangle$ shows periodic behavior (see
top panel of Fig.7) with short-range order corresponding to the energy maxima at quarter of the extended AB-period.
This may occur due to the loss of phase coherence between the bound pairs and
subsequent diminished superconductivity at those flux values. However, the
long-range order persists even at large time regime, showing the existence of
superconducting phase. On the contrary, in presence of strong electric field,
the $\langle P(r)\rangle$ approaches to zero at higher $\phi(t)=FLt$ values (see bottom panel of Fig.7),
indicating the breakage of Cooper-pairs. This proves the superconducting to
metallic phase transition. Note that, the value of $\langle P(r)\rangle$ does
not go to its ideal value of zero due to probable finite size effect.

%******************************** Fig.8 *****************************
\begin{figure}[ht]
\begin{center}
\includegraphics[width=5.0in]{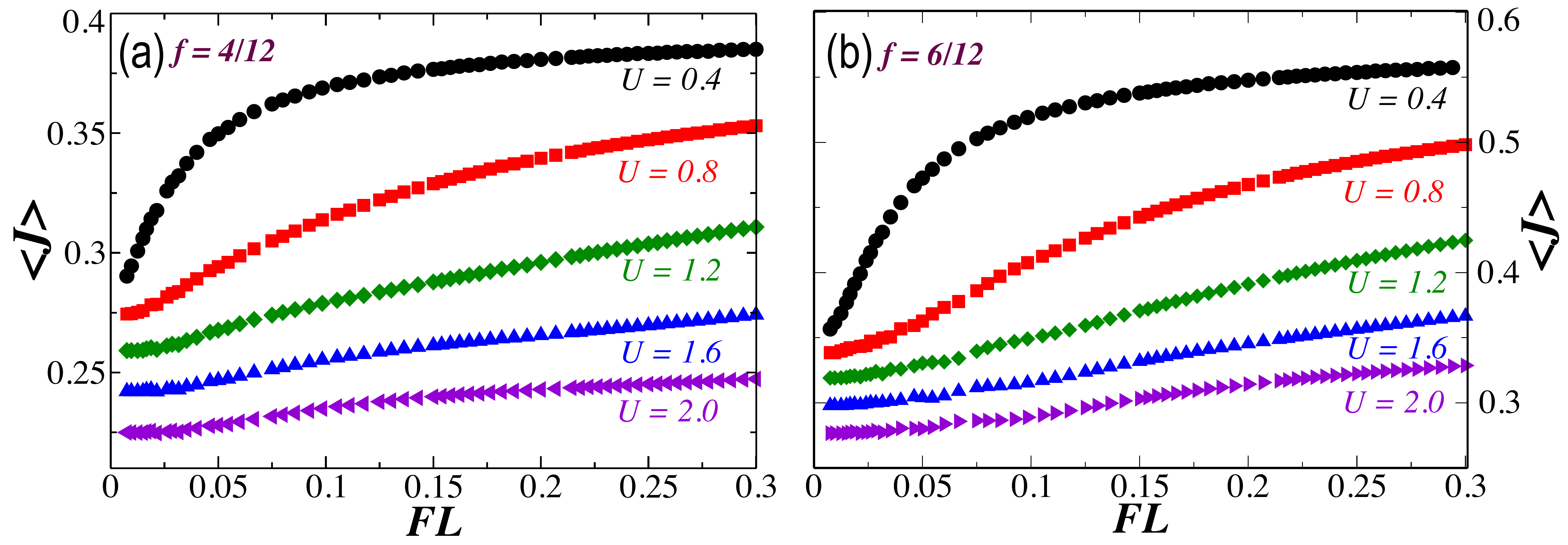}
\caption{Plot of time-averaged current, $\langle J \rangle$ as a function
of $FL$ for different values of attractive interaction $U$ with filling factors (a) $f=4/12$ and (b) $f=6/12$.}
\end{center}
\end{figure}
%********************************************************************

To gain further insight, we calculate the time-averaged current as follows\cite{takashi},

\begin{equation}
\langle J \rangle=\frac{1}{T} \int^T_0 \langle J(t)\rangle dt
\end{equation}

\noindent with integration over quarter of the extended AB period ($\phi(t) = L/4$)
and plot that as a function of $FL$ in Fig.8. As we have discussed before, the
increase in applied electric field breaks the electron-pairs and that can cause
the increase in the induced current. As can be seen from Fig.8(a) and (b), at lower values
of $FL$, the induced current does not increase initially. However, beyond a
certain critical strength of $FL$, the $\langle J \rangle$ starts increasing.
The critical value of $FL$ increases with the increase in the strength of the
attractive potential, $U$.
That necessitates the application of stronger $FL$ to trigger
the superconducting to metallic phase transition. Note that, at higher $FL$
regime, the $\langle J \rangle$ gets saturated mainly depending on the available
conduction electrons. All these observations are consistent for different filling factors.

\section{Conclusion}
In conclusion, we have studied the time-dependent non-equilibrium properties
of quasi-one-dimensional superconducting ring under the influence of external
electric field and the corresponding flux quantizations of energy and current density.
We present the numerical results of the systems with varying length and filling factor.
At half-filling, the formation of the charge density wave ground state restricts the 
AB-period within the lattice constant. Hole doping leads to 
superconducting ground state with halved extended AB-period, although with fractional
peaks, owing to the level anticrossings of degenerate states. Further increase in
hole doping reduces the degeneracy and fades away the fractional periodicity.
We observe that, the applied field breaks the electron-pairs, namely the Cooper-pairs 
in these rings, driving the system from a superconducting to a metallic phase. The 
required strength of this applied field depends on the strength of the attractive 
interaction potential. Our study on the non-equilibrium behavior of the superconducting 
rings will drive further experiments to explore the rich phase diagram of the strongly 
correlated low-dimensional systems under external perturbations.

\section{Acknowledgment}
B.P. thanks the UGC, Govt. of India for support through a fellowship and
S.K.P. acknowledges DST, Govt. of India for financial support.
S.D. acknowledges the ICYS-MANA, NIMS, Japan where part of this work has been done.\\\\

\end{document}